%% file: main.tex
\documentclass[aps,prl,reprint,superscriptaddress,amsmath,amssymb]{revtex4-2}

\usepackage{graphicx}
\usepackage{bm}
\usepackage[dvipsnames]{xcolor}
\usepackage{ulem}
\usepackage{subcaption}
\usepackage{siunitx}
\usepackage{enumitem}
\setlist{nolistsep}
\newlength{\mainparindent}
\usepackage{hyperref}
\hypersetup{colorlinks=true,linkcolor=blue,citecolor=blue,urlcolor=blue}
\usepackage{placeins}

\begin{document}

\title{Bimodal colloids highlight the structural mirror of rigidity percolation and yielding}

\author{Robert A. Campbell}
\affiliation{Department of Mechanical and Industrial Engineering, Northeastern University, Boston, MA 02115, USA}

\author{Ziye Zhuang}
\affiliation{Department of Chemical and Biomolecular Engineering, University of California, Irvine, CA 92697, USA}

\author{Ali Mohraz}
\affiliation{Department of Chemical and Biomolecular Engineering, University of California, Irvine, CA 92697, USA}

\author{Safa Jamali}
\email{s.jamali@northeastern.edu}
\affiliation{Department of Mechanical and Industrial Engineering, Northeastern University, Boston, MA 02115, USA}

\date{\today}

\begin{abstract}
In metastable particulate gels, it is tempting to believe that the dynamic similarities between the fluid-to-solid non-linear phase transition of rigidity percolation and the solid-to-fluid transition that occurs during yielding represent mirror images of the same continuous process. Even though these behaviors are clearly dynamically similar, their multi-scale nature makes it difficult to determine if they could also follow a unified structural pathway. We know from model monodisperse colloidal gels that both yielding and the elastic modulus seem to be heavily influenced by a small subset of topologically distinct singly-connected bridges linking mesoscale features. Here we use particle simulations to examine the participation of different classes of particle-level bonds and their contributions to the bulk mechanical response. We find that rigidity is disproportionately supported by singly connected intercluster bridges, whereas yielding localizes at bonds with high edge-betweenness centrality (EBC); strikingly, these independently identified populations substantially overlap and perform comparable mechanical roles. Bimodality exposes this correspondence by concentrating large-particle contacts in both populations, thereby providing a compositional label for the common backbone. Thus, rigidity and yielding are opposing mechanical manifestations of the same mesoscale structure: the intercluster bottlenecks that establish rigidity are also the sites at which rigidity is preferentially lost.
\end{abstract}

\maketitle

% =====================================================
% Main Text (2 column)
% =====================================================

The mechanical properties of amorphous particulate systems are strongly influenced by their underlying network structure \cite{Zaccarelli2007,Rocklin2021,Zia2014,Varga2015,Varga2015_hydro,Sherman2020}. A notable example is colloidal gels, where a low volume fraction of attractive particles will aggregate into a metastable, mechanically rigid, space-spanning network. In these and related soft matter systems, the details of this underlying network influence the full range of viscoelastic behavior, from linear elasticity to non-liner stiffening, yielding, and flow \cite{Becu2006,Colombo2014,Johnson2019,RajaramMohraz2010,Hsiao2012}. The intrinsic structure-mechanics relationships that define this behavior have motivated significant research into the quantification and design of these networks \cite{Whitaker2019,Nabizadeh2024,Mangal2024,Smith2024,Rocklin2021, mangal2023topological}. However, it is difficult to get a detailed understanding of the mechanically relevant portions of these networks because they are disordered and highly hierarchical.

From the standpoint of structure and dynamics, the emergence of mechanical rigidity in these systems can be described as the percolation of locally rigid units through a continuous non-equilibrium phase transition \cite{Tsurusawa2023,Rouwhorst2020}. This percolation has been shown to be a multi-scale effect, rather than a direct outcome of local particle connectivity \cite{PanizHaghighi2025,Dias2025}. The resulting colloidal gel structure can also be modeled as an assembly of cluster-level features. When these systems are viewed as a collection of locally isostatic cluster "blobs" it is possible to recover their elastic modulus using either the Cauchy-Born theory \cite{ZacconeWuDelGado2009,Whitaker2019,Nabizadeh2024} or a recursive rheological ladder model of viscoelastic elements \cite{Bantawa2023}. In both cases, mesoscale features are identified geometrically from the network, either using a correlation length \cite{Whitaker2019,Bantawa2023} or through graph theoretic clustering approaches \cite{Nabizadeh2024}. Similar network-based analysis methods have also been successful in describing the mechanics of dense, granular materials based on community level details of their contact networks \cite{Bassett2015,Papadopoulos2016,Papadopoulos2018}. Together, these results point to the significance of mesoscale connectivity in global rigidity percolation and the emergence of a bulk elastic response. In particular, the recursive rheological ladder model emphasizes the importance of singly-connected bonds that bridge more densely connected regions of the network, which effectively define the mechanical components of the overall structure. However, in the absence of bending rigidity or other rotational constraints, colloidal systems form comparatively dense strands with an average coordination number around six, where singly connected contacts are rare\cite{Tsurusawa2019,Zhang2019,Zhuang2026}. While the degree of coordination can be tuned in experimental systems \cite{Zhuang2026}, in high-coordination cases it is less clear which bonds define the mechanical response. This is particularly true in highly polydisperse systems or systems with bimodal populations of different sized particles, where size-disparities can produce enhanced geometric packing that influences the formation of rigid tetrahedral units \cite{WaheibiHsiao2024,LonialWeeks2026} and may alter the cluster-level landscape further.

Conversely, the breakdown of mechanical elasticity during the yielding transition has been shown to be another non-equilibrium percolation transition in colloidal glasses \cite{Ghosh2017,Shrivastav2016}, prompting some to suggest that rigidity percolation and yielding may function as two sides of the same continuous process \cite{Rouwhorst2020}. It is tempting to then believe that the structural features underlying both fluid-to-solid and solid-to-fluid transitions should mirror each other, although this has not been rigorously tested. The yielding and fracture of soft amorphous materials is increasingly understood to function not as a bulk bond rupture and fluidization process, but as the accumulation of local microplastic rearrangements during periods of solid-like deformation \cite{Grenard2014,Leocmach2014,Cho2022,Divoux2024}. Simulations have shown that these microplastic events are elastically driven and can rearrange within the network, as stress released by one event is redistributed, triggering subsequent microplastic events in other locations \cite{Bouzid2017}. In colloidal glasses, the correlation and directed percolation of these events precedes macroscopic failure \cite{Ghosh2017,Shrivastav2016}; however, it is not clear if this multi-scale phenomenon has a comparable mesoscale analog to the cluster "blobs" in mechanical rigidity. Some evidence for mesoscale mediated yielding exists in glasses and gels that exhibit two step yielding, where the first step has been shown to correspond to the breaking of interparticle bonds, and the second step corresponds to the fragmentation of clusters at a secondary characteristic length-scale \cite{Koumakis2011,Moghimi2020}. Additionally, in both highly fractal systems and in central-force dominated depletion systems, experimental work has shown that colloidal gel yielding is preceded by local bending of singly-connected soft pivots \cite{Mohraz2005,RajaramMohraz2010}. These soft pivots are phenomenologically similar to the singly-connected bridges relevant to elastic mechanics; however, it is not clear whether the same structural feature can perform both pivot and bridge functions in the network, or the extent to which pivots exist as a distinct class of microplastic events.

Here we unify the structural understanding of these two non-equilibrium processes using graph theoretic network analysis. Geodesic centrality, which identifies features that are central to the bulk connectivity of the network, has successfully identified mechanically relevant features in both dense suspensions \cite{Papadopoulos2018,KollmerDaniels2019,NabizadehPRL2022} and colloidal gels \cite{Nabizadeh2024,mangal2023topological}. In colloidal gel contact networks, it has been shown that the bonds with the highest edge-betweenness centrality (EBC) values at the quiescent state preferentially break over the course of the yielding transition \cite{Mangal2024}. Although it is somewhat surprising that the quiescent state can predict a nonlinear mechanical response, the relative significance of these bonds as compared to the rearrangement and formation of new features is reminiscent of experimental studies showing the mechanical aging of a system at rest through the stiffening of existing critical bonds \cite{Bonacci2020} and the retention of memory of an initial state under isotropic compression \cite{Milani2026}. In this study, we isolate the subset of bonds with the top 10\% of EBC values in simulations of both monomodal and bimodal (size ratio 1:2) colloidal depletion gels. We also use topological analyses to identify locally rigid simple tetrahedra \cite{LonialWeeks2026}, and inter-cluster bridges \cite{Nabizadeh2024,Zhuang2026}, comparing these three populations of bonds at the quiescent state and then monitoring their behavior under simple shear. This allows us to track the structural and mechanical roles of each population. We find that not only is there significant overlap between bridges and high EBC bonds, but both populations perform similar mechanical functions, suggesting that bridges and soft pivots are the same contacts. The high accuracy of EBC in identifying these structures also effectively allows us to identify mesoscale features from a bulk structural measure. We exploit bimodality as a structural contrast: because large particles support more and stronger contacts, their placement amplifies otherwise sparse mechanically consequential features. If rigidity-bearing bridges and yielding-prone high-EBC bonds are manifestations of a common mesoscale backbone, the two independent classifications should select the same large-particle-rich contacts.

\begin{figure*} [th!]
    \centering
    \includegraphics[width=1\linewidth]{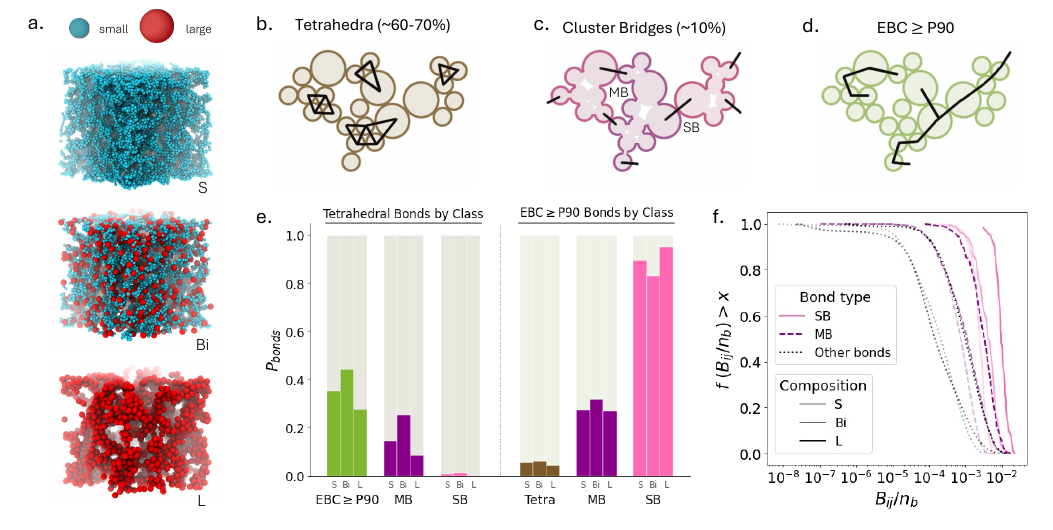}
    \caption{Comparison of structural features in monomodal and bimodal colloidal gels. (a) Visualization of the three systems under study, small particle monomodal gel, bimodal gel, and large particle monomodal gel. (b) A 2D schematic of a region of the bimodal gel labeled with the bonds that participate in rigid tetrahedral units, which make up roughly 60\% of all bonds. (c) The same schematic region showing inter-cluster bridges, which make up roughly 10\% of all bonds and can be divided into two classes: multi-connected bridges (MB), where multiple parallel connections bridge two clusters, and singly-connected bridges (SB), where only one contact connects two clusters. (d) The same schematic region again, showing bonds with the top 10\% of EBC values. (e) Comparison of the fraction of high EBC (green), MB (purple), and SB (pink) bonds that participate in tetrahedra (left) vs. the fraction of tetrahedral (brown), MB (purple), and SB (pink) bonds that participate in high EBC edges (right). (f) A complementary cumulative distribution function (CCDF) plot showing the fraction of bonds in a given population whose EBC value $B_{ij}$ is greater than the value at x. These values are decomposed by bond population into SB (pink, solid), MB (purple, dashed), and other (grey, dotted) bonds for all systems. All EBC values are normalized by the total number of bonds in the system, $n_b$, to ensure direct comparability between different sized networks.}
    \label{fig:bond-comp}
\end{figure*}

We compare the structure and mechanics of three colloidal gels: bimodal ($a_L/a_s=2, \phi_L/\phi=0.5$), monodisperse small particle, and monodisperse large particle. These systems were simulated using a Core-Modified Dissipative Particle Dynamics (CM-DPD) method, \cite{Whittle2010,Jamali2015,Boromand2017,Nabizadeh2021} implemented in a custom version of the open-source molecular dynamics toolkit HOOMD-blue \cite{Anderson2020}. The inclusion of a discrete solvent in this method ensures that long-range hydrodynamic interactions are preserved. Depletion interactions are implemented using a size-weighted Morse potential to account for the changes in volumetric exclusion between small and large particles. A representative 3D snapshot of each system is shown in Figure \ref{fig:bond-comp}a. From these systems, we construct a contact network of all colloidal pairs that fall within the physical interaction limit ($h_{ij}\leq \Delta$), representing the most simply connected network for this collection of attractive particles \cite{Nabizadeh2024,PanizHaghighi2025}. We then classify these network bonds into three topological classes: rigid tetrahedra, cluster-cluster bridges, and the top 10\% of EBC values (see Supplementary Material for additional methods details).

The schematic illustrations of these topological classes, shown in Fig\ref{fig:bond-comp}b-d, demonstrate that although individual tetrahedral structures are seen to aggregate together, they fully percolate through contacts in the more amorphous regions of the gel. The majority of inter-cluster bridges occur in these amorphous regions, and can be classified as either multi-connected bridges (MB) where multiple, parallel particle-particle bonds link two unique clusters, or as singly-connected bridges (SB), where only one particle bond links two unique clusters. The high EBC bonds do not necessarily fully percolate, but they do form large connected pathways that pass through both tetrahedral and bridge structures. Fig\ref{fig:bond-comp}e shows that roughly 40\% of tetrahedra participate in high EBC bonds, and enhanced geometric packing in the bimodal system does allow for more tetrahedra to serve as MB contacts; however, because of their dense structure very few tetrahedra serve as SB contacts. Conversely, tetrahedral bonds make up less than 10\% of the highest EBC bonds, and there is very high overlap between SB and high EBC bonds. The complementary cumulative distribution function (CCDF) plot in Fig\ref{fig:bond-comp}f confirms that both MB and SB contacts consistently have the highest EBC values, representing a distinct topological class in these networks.

\begin{figure}[ht!]
    \centering
    \includegraphics[width=1\linewidth]{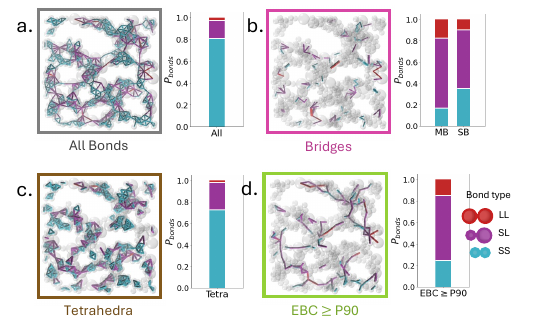}
    \caption{Bond compositions by particle type. A representative z-slice of the bimodal network is visualized four times for different segments of the network. Each one is color-coded by bond type, small-small (SS, teal), small-large (SL, purple), and large-large (LL, red). Each visualization is paired with a stacked bar plot showing the relative fraction of bond types in that segment of the network. (a) The entire network, (b) the inter-cluster bridges (SB in bold), (c) the tetrahedral structures, and (d) bonds with the top 10\% of EBC values.}
    \label{fig:bi-preference}
\end{figure}

Labeling each bond in the bimodal system by its component particle types, we find in Fig. \ref{fig:bi-preference} that the high EBC and cluster bridge structures show a distinct preference for large and size-anisotropic contacts. The concentration of large particles in these structures effectively labels these as mechanically relevant features. Despite making up less than 20\% of the total network (Fig \ref{fig:bi-preference}a), bonds that include large particles make up roughly 80\% of both the high EBC and bridge contacts (Fig \ref{fig:bi-preference}b,d). Since large particles can have more neighbors, they are less represented in SB bonds than MB bonds, but they are still significantly more prevalent in these bridges than the internal cluster bonds. In contrast, tetrahedral structures follow the bulk composition more closely (Fig \ref{fig:bi-preference}c), and visualizing these different components highlights the way that tetrahedral packing relies on compact, small-small bonds, frequently excluding bridges. The overlap between bridges and high EBC bonds, with each class including roughly 35\% of the other, produces what appears to be an alternate mechanically stabilizing structure throughout the network. 

\begin{figure} [ht]
    \centering
    \includegraphics[width=1\linewidth]{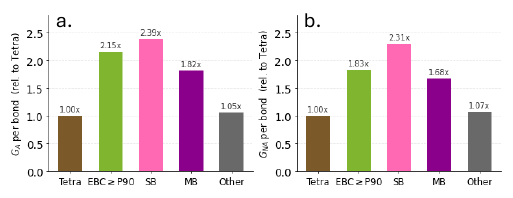}
    \caption{Relative contributions to the static shear modulus of the bimodal system. (a) Decomposition of the affine (Born) contribution to the static shear modulus by bond type: rigid tetrahedra, high EBC bonds, singly-connected cluster bridges, multi-connected cluster bridges, and other bonds in the network. All values are normalized by the contribution from tetrahedral bonds. (c) The same decomposition for the non-affine contribution.}
    \label{fig:mechanics}
\end{figure}

To evaluate the mechanical significance of these different classes, we calculate their relative contributions to the static shear modulus following the athermal decomposition of affine and non-affine contributions for model disordered solids, where $G=G_{A}-G_{NA}$ \cite{LemaitreMaloney2006,ZacconeScossaRomano2011,Zaccone2023,Nabizadeh2024}. The affine (Born) contribution is calculated from the pair-specific Morse potential, $U_{ij}$ as $G_A = (1/V) \Sigma_{bonds} (U_{ij}''-U'_{ij}/r_{ij}) r_{ij}^2 (n_x n_y)^2$. The non-affine contribution is calculated directly from an affine force field, calculated per bond as $\Xi_{i}=-\Sigma_j (U_{ij}''-U'_{ij}/r_{ij})r_{ij}n_xn_y\hat{n}_{ij}$, and solving the linear response $Hu=\Xi$ on the full configuration-specific Hessian, such that $G_{NA} = (1/V) \ \Xi \cdot H^{-1} \cdot \Xi$.  In the bimodal system, the size-dependent depletion interaction creates a stiffness hierarchy between small-small, small-large, and large-large bonds, where stiffness $k_s\propto U''_{ij}$, $U_0^{SL}=1.5U_0^{SS}$, and $U_0^{LL}=2U_0^{SS}$. The stiffness hierarchy magnifies the mechanical importance of bonds that include large particles, which are preferentially concentrated at the mesoscale cluster bridges and high EBC contacts, allowing us to resolve the mechanically relevant mesoscale features from microscale interactions. Fig. \ref{fig:mechanics} shows that, when compared to the contribution of rigid tetrahedra, we see that bridges, especially SB contacts, contribute roughly twice as much to both $G_A$ and $G_{NA}$. This is in agreement with hierarchical models that estimate bulk rigidity from the mesoscale contacts between locally rigid clusters \cite{ZacconeWuDelGado2009,Whitaker2019,Nabizadeh2024,Bantawa2023}. The high EBC bonds contribute at similar rates despite only including around 35\% of the total bridges, likely because of their comparable composition of large particle bonds.

\begin{figure}[ht]
    \centering
    \includegraphics[width=1\linewidth]{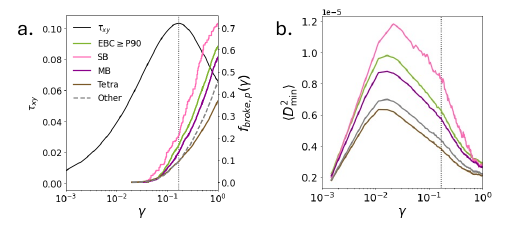}
    \caption{Relative contributions to yielding of the bimodal system. All data is decomposed by topological class: high EBC (green), SB (pink), MB (purple), tetrahedral (brown), and other (grey) bonds. The black dashed line marks the location of the stress overshoot. (a) The stress-strain curve (black) overlaid with the cumulative fraction of quiescent bond breaks. (b) Per-particle non-affine motion $D^2_{min}$.}
    \label{fig:yielding}
\end{figure}

We also monitor the behavior of these bonds during the yielding transition. We apply simple shear flow on the bimodal system using a shear rate, $\dot{\gamma}$, corresponding to a high dimensionless Mason number for small-small interactions of $Mn_{SS}=6\pi\eta\dot{\gamma}a_{S}^3/U^{SS}_{0}=5.9$, and track the breaking of individual bonds that were present in the quiescent state, following previously reported methods \cite{Nabizadeh2021,Mangal2024}. Fig \ref{fig:yielding}a shows the stress response overlaid with the cumulative fraction of broken bonds in each class. These results qualitatively match previous work \cite{Colombo2014} that showed colloidal gel bonds begin breaking slightly before the stress overshoot. At this high shear rate, we find that roughly 60-70\% of bridges and high EBC bonds break over the course of the simulation, compared with roughly 40\% of other bonds. We also calculate the per-particle $D^2_{min}$ non-affine displacement relative to a purely affine response \cite{FalkLanger1998}, shown in Fig. \ref{fig:yielding}d. This reveals that both bridges and high EBC bonds see roughly 2-3x larger non-affine deformation than the other particles in the network, suggesting that they dominate the microplastic events that initiate yielding. 

The unique mechanical and topological roles of inter-cluster bridges and high EBC bonds present a potential pathway for future extension of the recursive rheological ladder model \cite{Bantawa2023} to more densely coordinated particulate systems. 

These results show that mesoscale bridge models successfully identify mechanically important connections between clusters, but these methods still require explicit partitioning of the disordered network. While we find that singly-connected contacts have notable mechanical significance, even when they make up less than 1\% of the total system, they are supported in this role by a secondary set of topologically similar structures. We find that global edge-betweenness centrality identifies a mechanically equivalent backbone directly from the unpartitioned contact graph. Furthermore, bimodality makes this backbone compositionally visible because large particles preferentially occupy high-centrality inter-cluster contacts. In the bimodal gel, large-particle contacts independently mark both bridge-defined rigidity and high-EBC yielding sites, demonstrating that the correspondence is structural rather than merely dynamical. Rigidity and yielding therefore emerge as opposing mechanical manifestations of a common mesoscale backbone: the intercluster bottlenecks that transmit load are also the sites that concentrate nonaffine deformation and initiate failure. Particle-size bimodality makes this structural mirror directly visible.

\begin{acknowledgments}
Financial support for this study was provided to S.J. and A.M. by the National Science Foundation (PMP-2025613) and to R.A.C. and S.J. by NASA ROSES FINESST (80NSSC23K0015). Computational resources were provided by the Massachusetts Green High-Performance Computing Center in Holyoke, MA.
\end{acknowledgments}

% main text references
\input{mainbib.bbl}

% =====================================================
% Supplemental Material (single column)
% =====================================================
% match paragraph indent
\setlength{\mainparindent}{\parindent}

\clearpage
\onecolumngrid\linespread{1.3}\selectfont

% widen SI margins
\addtolength{\oddsidemargin}{1cm}
\addtolength{\evensidemargin}{1cm}
\addtolength{\textwidth}{-2cm}
\setlength{\hsize}{\textwidth}
\setlength{\columnwidth}{\textwidth}
\setlength{\linewidth}{\textwidth}

% Use main-text paragraph indent for the whole SI.
\setlength{\parindent}{\mainparindent}

\setcounter{section}{0}
\setcounter{subsection}{0}
\setcounter{figure}{0}
\setcounter{table}{0}
\setcounter{equation}{0}
\renewcommand{\thesection}{S\arabic{section}}
\renewcommand{\thesubsection}{S\arabic{subsection}}
\renewcommand{\thefigure}{S\arabic{figure}}
\renewcommand{\thetable}{S\arabic{table}}
\renewcommand{\theequation}{S\arabic{equation}}

% restyle sections to be left-justified
\makeatletter
\setcounter{secnumdepth}{0} 
\renewcommand\subsection{\@startsection{subsection}{2}
  {-\parindent}
  {2.5ex \@plus 1ex \@minus .2ex}
  {1.2ex \@plus .2ex}
  {\normalfont\normalsize\bfseries}}
% restyle subsubsections to be left-justified
\renewcommand\subsubsection{\@startsection{subsubsection}{3}
  {-\parindent}
  {2.0ex \@plus 1ex \@minus .2ex}
  {1.0ex \@plus .2ex}
  {\normalfont\normalsize\textit}}
\makeatother

% title
\begin{center}
{\Large\bfseries Supplemental Material for:\\[3pt]
Bimodal colloids highlight the structural mirror of\\ rigidity percolation and yielding\par}
\vspace{10pt}
Robert A. Campbell,$^{1}$ Ziye Zhuang,$^{2}$ Ali Mohraz,$^{2}$ and Safa Jamali$^{1,\ast}$\\[6pt]
{\itshape $^{1}$Department of Mechanical and Industrial Engineering,\\ Northeastern University, Boston, MA 02115, USA\par}
\vspace{2pt}
{\itshape $^{2}$Department of Chemical and Biomolecular Engineering,\\ University of California, Irvine, CA 92697, USA\par}
\end{center}
\vspace{1em}

\tableofcontents

% SI text
\newpage
\phantomsection
\subsection{Simulation Methods}

Simulation of colloidal gelation and yielding follows established Core-Modified Dissipative Particle Dynamics (CM-DPD) methods for explicit-solvent simulations of colloidal gels \cite{Whittle2010,Jamali2015,Boromand2017,Nabizadeh2021}. Particle size differences are included by assigning small colloids a radius $a_S=1$, and large colloids a radius $a_L=2$. They are density matched such that $m_C = 4/3 \rho \pi a_C^3$, where $\rho=3$ is the solvent number density. All systems are initialized in a cubic box of side length $L= 70a_S$ with Lees-Edwards periodic boundary conditions. The total volume fraction of colloids is held constant at $\phi = 0.2$, and the bimodal system is composed of a 50:50 volumetric split, such that $\phi_S=\phi_L=0.1$. 

Depletion attraction is implemented using a Morse potential \cite{Zia2014}. To account for the size-dependent nature of depletion as a volumetric exclusion effect, we scale the strength of attraction with the average size of each interacting colloid pair: $U_0^{ij}=D_0(a_{i}+a_{j})/2$. At particle size ratio 1:2, this approximation produces attraction strengths in agreement with the Asakura-Oosawa-Vrij model \cite{Asakura1958,Vrij1976}. All systems are gelled under the same depletion condition of $D_0 = 12k_BT$ and $\kappa{a_S}=60$, for an effective interaction distance of $0.05a_S$. Because this interaction range is quite short, we extend it to $\kappa a=10$ during yielding, for an interaction distance of $0.3a_s$. This approximation improves the resolution of microplastic events and produces more realistic stress-strain behavior, as demonstrated in previous parametric studies \cite{Mangal2024}. We qualitatively validate this choice by running yielding simulations at multiple interaction ranges to confirm that the quiescent topological features follow similar trends in all conditions. 

During gelation, network formation is confirmed by monitoring the ensemble-averaged coordination number, system percolation, and mean squared displacement over time, as in previous studies \cite{Nabizadeh2024,PanizHaghighi2025}. To ensure the existence of a stable network, the final gel structure is evaluated at roughly 20x the time to network percolation.

\subsection{Tetrahedra Identification}

Rigid tetrahedral units are defined as bonds that participate in 4-cliques (sets of six bonds that connect four adjacent vertices) in the network. This structure is an established simple and useful unit of local rigidity in colloidal gels \cite{LonialWeeks2026}. We also classify sets of seven fully connected colloids (7 nodes, 16 edges) as pentagonal bipyramids, a stricter definition of rigid units that can better distinguish between amorphous and crystalline pathways to rigidity percolation \cite{Tsurusawa2023}. However, we do not observe the formation of pentagonal bipyramids in our gels, consistent with recent experiments that emphasized the role of simple tetrahedra in the low-$\phi$ gel regime \cite{WaheibiHsiao2024,LonialWeeks2026}. 

\subsection{Gaussian Mixture Model Based Clustering}

Effective cluster structure is determined using an established graph-based Gaussian Mixture Model (GMM) method that groups particles together based on similar network topology \cite{Nabizadeh2024,Zhuang2026}. This method follows four steps which are described in full detail, below:

\begin{enumerate}
\item The 3-dimensional colloidal contact network is translated into a topological latent space with 128 dimensions
\item This network is reduced to a 10-dimensional embedding
\item Particles are grouped into clusters based on the topological properties of this 10D network
\item The resulting clusters are physically validated in the original 3D space
\end{enumerate}

Here we use an unweighted contact network, where nodes represent individual particles, edges represent attractive pairwise bonds through the model interaction potential, and all edges are given equal "weight" when determining network structure. This gives us a purely topological representation of the most simply connected network for a collection of attractive particles.

\subsubsection{1. Node Representation in Latent Space.} We apply the semi-supervised node2vec model\cite{Grover2016node2vec} to obtain a particle representation matrix in 128 dimensions. This model uses a stochastic gradient descent approach to learn the feature representation of nodes in a network. A neighborhood set is defined for every node in the network using a fixed-length second order random walk sampling strategy. This method is guided by the parameters $p$ and $q$, which control how fast the walk explores the neighborhood of the starting node (in this study we use the default values: $p=q=1$). The network is traversed for a given number of walks of a set length (in our case 20 walks of length 80 edges) to label the neighborhoods of each node. In the neighborhood of node, $t$, if a node $v$ is visited in the initial random walk from $t$, then the transition probability is set by the distance of $t$ to the neighbors of $v$:
\begin{equation}
\alpha_{pq}(t,x) =
    \begin{cases}
      \frac{1}{p} & \text{if } d_{tx}=0,\\
      1 & \text{if } d_{tx}=1,\\
      \frac{1}{q} & \text{if } d_{tx}=2,
    \end{cases}  
\label{eq:node2vec_neighborhood}
\end{equation}

This set of network neighborhoods is then used to maximize the log probability of observing a neighbor for a given node, conditioned by it's feature vector. This is performed using stochastic gradient descent in the objective function:
\begin{equation}
\genfrac{}{}{0pt}{}{\text{max}}{f} \ \ \sum_{u \in V} \biggl( - \log \bigg( \sum_{v \in V} e^{f(u) \cdot f(v)} \bigg) + \sum_{v \in N_S(u)} f(v) \cdot f(u) \biggl)
\label{eq:node2vec_grad-descent}
\end{equation}
\noindent
where $f$ is the set of feature representation vectors, $V$ is the set of nodes in the network, and $N_S(u)$ is a neighbor of node $u$.

\subsubsection{2. Dimensional Reduction.} Dimensional reduction of the 128D latent space network is performed with Uniform Manifold Approximation and Projection (UMAP)\cite{Mcinnes2020umap,McInnes2018}. First, a fuzzy simplicial complex with the Riemannian geometry theoretical framework is constructred from the 128D network. This step produces a weighted graph describing the manifold structure of the data. This weighted graph is passed to a secondary forced-directed graph layout that defines open sets of data points, and then assigns weights between each pair of overlapping open sets. Each open set is defined as an n-dimensional sphere, with the radius defined by a local distance function tuned to include $k$ nearest neighbors of a given data point. The edge weight between the data points $x_i$ and $x_j$ therefore becomes a function of the geodesic distance of each point on the Riemannian manifold of the other: $w_{x_ix_j}=a+b- a \times b $, where $a$ is the geodesic distance of $x_i$ on the Riemannian manifold of $x_j$, and $b$ is the geodesic distance of $x_j$ on the Reimannian manifold of $x_i$.

A set of attractive and repulsive forces is then iteratively applied to a sample of and edges to optimize the edgewise cross-entropy between the initial weighted graph and an equivalent weighted graph constructed from new points $y_i$ and $y_j$ embedded in the target dimension. Here each $y$ is initialized by the eigenvector of the normalized Laplacian matrix of the fuzzy graph constructed in the first step. Equation \ref{eq:UMAP-att}defines the attractive force used in this optimization step, and Equation \ref{eq:UMAP-rep} defines the repulsive force.

\begin{equation}
\frac{-2ab||y_i-y_j||_2^{2(b-1)}}{1+||y_i-y_j||_2^2} w_{x_i,x_j}(y_i-y_j)
\label{eq:UMAP-att}
\end{equation}

\begin{equation}
\frac{2b}{\big( \epsilon + ||y_i-y_j||_2^2 \big) \big( 1+a||y_i-y_j||_2^{2b} \big)} (1-w_{x_i,x_j})(y_i-y_j)
\label{eq:UMAP-rep}
\end{equation}

Here we use UMAP to reduce our network into a 10-dimensional space for clustering. This method can also be used to reduce the system back to 3D space for direct comparison with the original particle configuration.

\subsubsection{3. Gaussian Mixture Model (GMM) Particle Clustering.} GMM is a widely used mixture model that assumes each base distribution has the form of an unknown multivariate Gaussian distribution. For $k$ distributions, each distribution is given the mixing coefficient $\pi_i$, where $\pi_i,\forall i \in {i,...,k}$ indicates the contribution of each model to the overall distribution, such that $0 \leq \pi_i \leq 1$ and $\sum_{i=1}^k \pi_i = 1$. The marginal distribution of a point $x_n$ can therefore be written as a Gaussian distribution of the form:

\begin{equation}
p(x_n) = \sum_{i=1}^{k} \pi_i \mathcal{N}(x_n|\mu_i,\Sigma_i)
\label{eq:GMM-gaussian}
\end{equation}

The conditional probability that a cluster $i$ is responsible for explaining the data point $x_n$ is therefore calculated as the responsibility:

\begin{equation}
\gamma(z_{ni}) = \frac{p(z_i=1)p(x_n|z_i=1)}{\Sigma^k_{j=1}p(z_j=1)p(x_n|z_j=1)} = \frac{\pi_i\mathcal{N}(x_n|\mu_i,\Sigma_i)}{\Sigma^k_{j=1}\pi_j\mathcal{N}(x_n|\mu_j,\Sigma_j)}
\label{eq:GMM-responsibility}
\end{equation}
\noindent
where $z_i$ is the latent membership indicator for a point belonging to $i$, $p(z_i = 1)$ is the prior probability that a point comes from $i$, $p(x_n | z_i = 1)$ is the likelihood of observing point $x_n$ in cluster $i$, and $\mathcal{N}(x_n | \mu_i, \Sigma_i) $ is the base distribution evaluated at $x_n$ with mean $\mu_i$ and covariance $\Sigma_i$.

From the responsibility, an expectation-maximization algorithm is used to fit the mixture of $k$ Gaussians to the physical set of $N$ colloidal nodes back in 3D space. This is done by maximizing the objective function:

\begin{equation}
L(\mu,\Sigma,\pi) = \sum_{n=1}^N \ln\bigg( \sum_{i=1}^k \pi_i \mathcal{N}(x_n|\mu_i,\Sigma_i)\bigg)
\label{eq:GMM-objective-function}
\end{equation}

First, we evaluate $L(\mu,\Sigma,\pi)$ from an initial set of values: $\mu_i$, $\Sigma_i$, and $\pi_i,\forall i \in {i,...,k}$. Then we set the objective function to zero and re-estimate the parameters using $m_i=\sum_{n=1}^N\gamma(z_{ni})$ nodes assigned to cluster $i$, mean $\mu_i = \frac{1}{m_i}\sum_{n=1}^N \gamma(z_{ni})$, covariance $\Sigma_i= \frac{1}{m_i}\sum_{n=1}^N \gamma(z_{ni})(x_{n}-\mu_i)(x_n-\mu_i)^T$, and mixing coefficient $\pi_i = \frac{m_i}{N}$. This step is then repeated until the values that allow the objective function to meet the convergence criteria are found.

Finally, we identify the optimal number of clusters by running GMM across a wide range of $k$ values and selecting the one that minimizes the Bayesian Information Criterion (BIC):

\begin{equation}
BIC(k,\mu,\Sigma,\pi)=k\ln(N)-2 \ \times \ L^{\star}(\mu,\Sigma,\pi)
\label{eq:GMM-BIC}
\end{equation}
\noindent
where $L^{\star}(\mu,\Sigma,\pi)$ is the is Equation \ref{eq:GMM-objective-function} for the GMM result corresponding to $k$ clusters.

\subsubsection{4. Cluster Validation in Real Space.} After assigning all particles to unique clusters using GMM, we validate the physical identity of these clusters. In this study, all of our original networks were fully connected (all particles belong to a single largest connected component), so we expect each cluster to also be a fully-connected subgraph. However, GMM is performed in topological latent space, and does not use the physical location of particles in 3D space when defining clusters. Therefore, it is possible that two topologically similar regions of the network that are far apart from each other will be grouped into a single cluster. In practice this is not common in our systems, but we still implement a physicality check to correct any assignment errors.

If a cluster subgraph is not fully connected, we evaluate the size of it's disconnected components. If the largest components makes up $\geq 90$\% of the total cluster, then we reassign the remaining disconnected components to their nearest physical neighbor. Otherwise we split the cluster into it's constituent components. If these components are $\geq 10$\% of the original cluster size, then we define them as new clusters. If these components are smaller than this, then we reassign them to their nearest neighboring physically connected cluster. 

\subsection{Edge-betweenness Centrality (EBC)}

Edge betweenness centrality (EBC) is the sum of the fraction of all shortest paths between two nodes in a network. It effectively identifies the extent to which an edge functions not only as a connection between two contacts, but also serves a central role in the bulk connectivity of all possible node pairs. It was first proposed as part of the Girvan-Newman algorithm \cite{GirvanNewman2002} for identifying cohesive communities by dropping the most central edges in a network. The EBC value, $B_{ij}$, of a given contact is calculated as:

\begin{equation}
B_{ij} = \sum_{a \neq b} \frac{\sigma(a,b|e_{ij})}{\sigma(a,b)}
\label{eq:doublelayer}
\end{equation}

where $\sigma(a,b)$ is the total number of shortest paths connecting all possible $(a,b)$ node pairs, and $\sigma(a,b|e_{ij})$ is the number of shortest paths that include the current edge, $e_{ij}$.

\subsection{Additional Yielding Data}

\begin{figure}[ht!]
\centering
\includegraphics[width=0.8\linewidth]{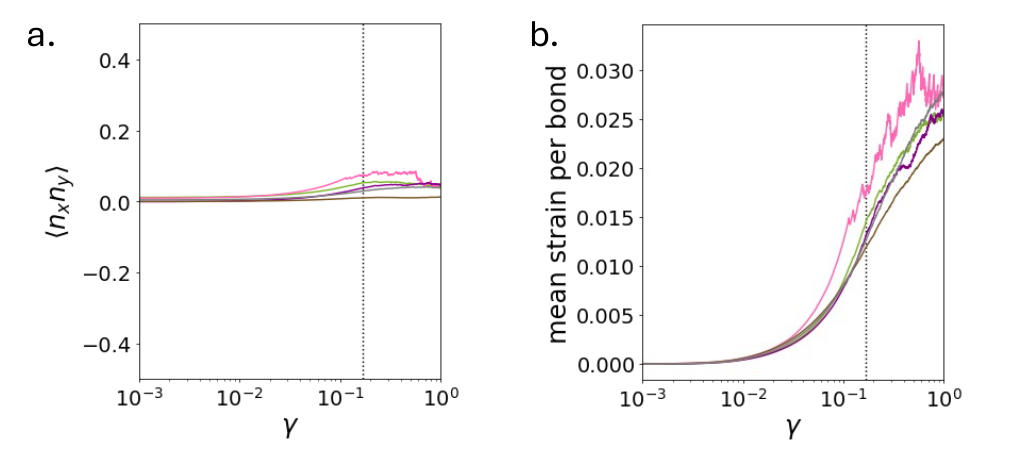}
\caption{Additional yielding measures. (a) Bond orientation in the shear direction, as calculated from the fabric tensor. (b) Cumulative mean strain per bond.}
\label{fig:SI_yield}
\end{figure}

In addition to the bulk stress, bond breaks, and per-particle non-affine deformation values reported in Fig. \ref{fig:yielding} we calculate the fabric tensor to measure the change in bond-orientation with respect to the shear direction and the mean strain per bond during yielding. Fig. \ref{fig:SI_yield}a,b shows that these measures have comparable values for all topological classes in the network, although rigid tetrahedra so the least amount of reorientation in the strain direction, as expected for stable rigid features. Additionally, variation in these measures does not predict the yielding transition, and instead corresponds with the onset of bond breaking closer to the stress overshoot.

\FloatBarrier
% SI references
\input{sibib.bbl}

\end{document}